# Characterization of the reorientation transition in classical Heisenberg models with dipole interaction

A. Hucht and K. D. Usadel

Theoretische Tieftemperaturphysik and Sonderforschungsbereich 166,
Gerhard-Mercator-Universität, 47048 Duisburg, Germany

**Abstract**

In this paper the thermodynamics of thin ferromagnetic films are studied in the framework of the classical Heisenberg model with uniaxial anisotropy and long–range dipole interaction. The dependence of the order of the reorientation transition in these systems on the number of layers and on the distribution of anisotropies is investigated using both mean field theory and Monte–Carlo simulations.

Alfred Hucht

Theoretische Tieftemperaturphysik, Gerhard-Mercator-Universität

Lotharstraße 1, 47048 Duisburg, Germany

Tel.: x49-203-379-3486

Fax: x49-203-379-2965

EMail: fred@thp.Uni-Duisburg.DE

Thin magnetic films with a ground state magnetization perpendicular to the surface can undergo a phase transition into a phase with in-plane magnetization at a temperature $T_R < T_C$, the Curie temperature of the film, due to the competition of dipole interaction and surface anisotropy [1]. This reorientation transition has already been investigated theoretically in the framework of the two–dimensional anisotropic Heisenberg model with dipole interaction using classical [2] and quantum mechanical [3] mean field theory, renormalization group theory [4], spin wave theory [5], and Monte–Carlo simulations [6, 7]. While all calculations for a monolayer except [6] show a first order reorientation transition, the calculations in [3] show a second order transition for a system with $L = 3$ layers and vanishing anisotropy in the middle layer. In a recent paper on the order of the reorientation transition Moschel *et al.* [8] proposed that the order depends on the distribution of anisotropies through the layer.

In this paper we present ground state aspects and detailed phase diagrams obtained from both mean field theory and Monte–Carlo simulations in a classical anisotropic Heisenberg model with long range dipole interactions. The system consists of $L$ two–dimensional layers and it is described by the Hamiltonian

$$\mathcal{H} = -\frac{J}{2} \sum_{\langle ij \rangle} \mathbf{s}_i \cdot \mathbf{s}_j - \sum_i K_{\lambda_i} (s_i^z)^2 + \frac{D}{2} \sum_{ij} r_{ij}^{-3} \mathbf{s}_i \cdot \mathbf{s}_j - 3 r_{ij}^{-5} (\mathbf{s}_i \cdot \mathbf{r}_{ij})(\mathbf{r}_{ij} \cdot \mathbf{s}_j),$$

where the $\mathbf{s}_i = (s_i^x, s_i^y, s_i^z)$ are classical Heisenberg spins at position $\mathbf{r}_i = (r_i^x, r_i^y, r_i^z)$ on a simple cubic lattice and $\mathbf{r}_{ij} = \mathbf{r}_i - \mathbf{r}_j$. $J$ is the nearest-neighbour exchange coupling constant, $K_\lambda$ is the uniaxial anisotropy in layer $\lambda = 1 \ldots L$, and $D = \mu_0 g^2 \mu_B^2 / 4\pi a^3$ is the strength of the long range dipole interaction on a lattice with lattice constant $a$. We consider ferromagnetic coupled systems with $J = 1$ in this paper.

One open question concerning this model is the occurence of stripe do-mains in the ground state of the monolayer ($L = 1$). Taylor *et al.* [2] showed that for certain values of the dipole interaction $D$ and large anisotropy $K_1$ there are stripe domains with all spins pointing into the $\pm z$-direction, but they were unable to extend their calculations into the range where $D$ is small. Furthermore they proposed that for small $D$ the two–dimensional system should be in a ferromagnetic state, ie. the stripe domain width $N_{\text{GS}}^z$ should diverge for a finite value of $D$. This uncertain result called for further investigation.

We analytically calculated the ground state stripe domain width $N_{\text{GS}}^z$ and the ground state energy $E_{\text{GS}}^z$, with the result

$$N_{\text{GS}}^z = N_0 e^{\frac{J}{2D}} + O\left(e^{-\frac{J}{2D}}\right)$$

and

$$E_{\text{GS}}^z = -K_1 - 2J + D\left(\frac{\Theta_0}{2} - \frac{4}{N_0}e^{-\frac{J}{2D}}\right) + O\left(e^{-\frac{3J}{2D}}\right),$$

with the constants $N_0 = 0.871026$ and the two-dimensional dipole sum [9] $\Theta_0 = 9.0336$. Note that $N_{\text{GS}}^z$ is finite for all finite values of $D$, although it grows exponentially fast when $J/2D$ gets large.

In the limit of large $N_{\text{GS}}^z$ the ground state properties of the system can be determined exactly: We find that a system with only one layer ($L = 1$) always has a first order phase transition with a free energy independent of the orientation of the spins when $K_1 = 6.77522D$. When $L = 2$, the transition is first order only when the anisotropies in both layers are equal ($K_1 = K_2 = 6.52962D$), while it is second order when $K_1$ and $K_2$ are different. For $L = 3$ the situation is more difficult, as we only get a first order transition when $K_1 = K_3 = 6.52920D$ and $K_2 = 6.28402D$. These ground state results indicate that the occurence of a first order transition is a rather special case in this model.

We investigated the finite temperature regime with mean field theory and found finite regions in parameter space where the system undergoes a first order reorientation transition. In figure 1 the phase diagram for $L = 2$ layers, $J = 1$, $D = 0.017$ and $K_1 + K_2 = 0.25$ is depicted. At the phase boundaries the in-plane-component ($xy$) and the $z$-component ($z$) of the total magnetization become unstable. The lines of second order transitions are dashed and the lines of first order transitions are solid. The arrows shall symbolize the orientation of the spins.

In the Monte–Carlo simulations we considered systems with $N \times N \times L$ spins and periodic boundary conditions in the in-plane directions. The precise method of the simulation is described in [7]. The results of the Monte–Carlo simulations are in very good agreement with the mean field calculations: For $L = 1$ we only find first order reorientation transitions, while for $L = 2$ the transition is of second order when $K_1$ differs enough from $K_2$. To illustrate these results, in figure 2 the histogram of the polar angle $\vartheta$ of the total-magnetization vector in layer 1, $h(\vartheta_1)$, is depicted. When the transition is of second order ($\circ$), the distribution has one maximum around $\langle \vartheta_1 \rangle$, while at a first order transition ($\diamond$) the distribution has two maxima, one at $\vartheta_1 = 0$ and a second at $\vartheta_1 = \pi/2$ due to hysteresis effects. The differing results in [6], where a second order transition is found for $L = 1$, are very likely due to short simulation times and an insufficient evaluation method. While in [6] a typical run involved $4 \times 10^5$ MC sweeps, we performed $2 \times 10^6$ sweeps and additionally used a global spin flip, which reduced the autocorrelation time by a factor of approx. 200. A detailed account of our work will be published elsewhere.

*Acknowledgement:* One of the authors (A. H.) would like to thank A. Moschel for fruitful discussions. This work was supported by the Deutsche Forschungsgemeinschaft through Sonderforschungsbereich 166.

# Figure captions

Figure 1: Mean field phase diagram for $L = 2$ layers, $J = 1$, $D = 0.017$ and $K_1 + K_2 = 0.25$. At the phase boundaries the in-plane-component $(xy)$ and the $z$-component $(z)$ of the total magnetization become unstable. The lines of second order transitions are dashed and the lines of first order transitions are solid. The arrows shall symbolize the orientation of the spins.

Figure 2: Histogram $h(\vartheta_1)$ of the polar angle of the total magnetization vector in layer 1, as obtained from MC simulations, for $32 \times 32 \times 2$ spins. $J = 1$, $D = 0.017$, and $K_1 + K_2 = 0.25$, with $K_1 = 0.05 (\circ)$ and $K_1 = 0.125 (\diamond)$. The lines are guides to the eyes.

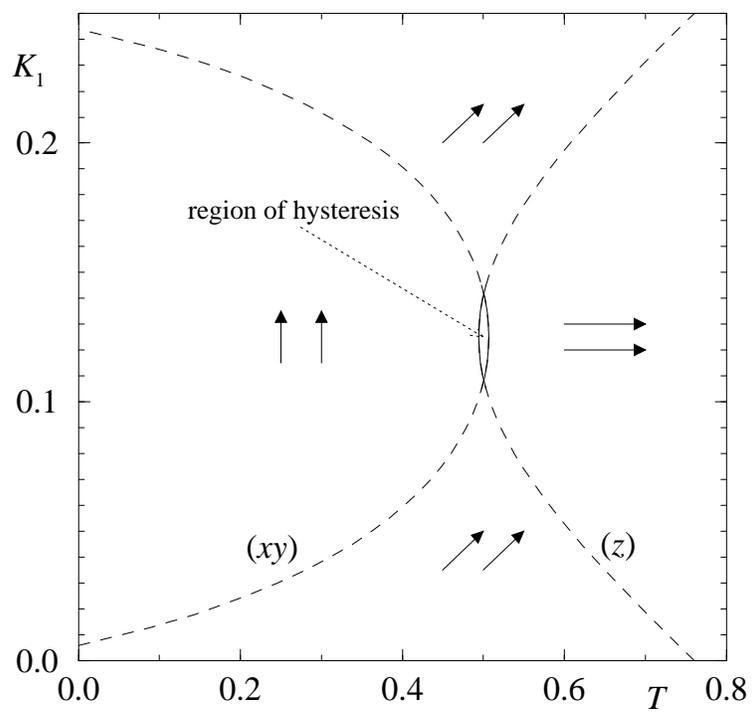

Figure 1

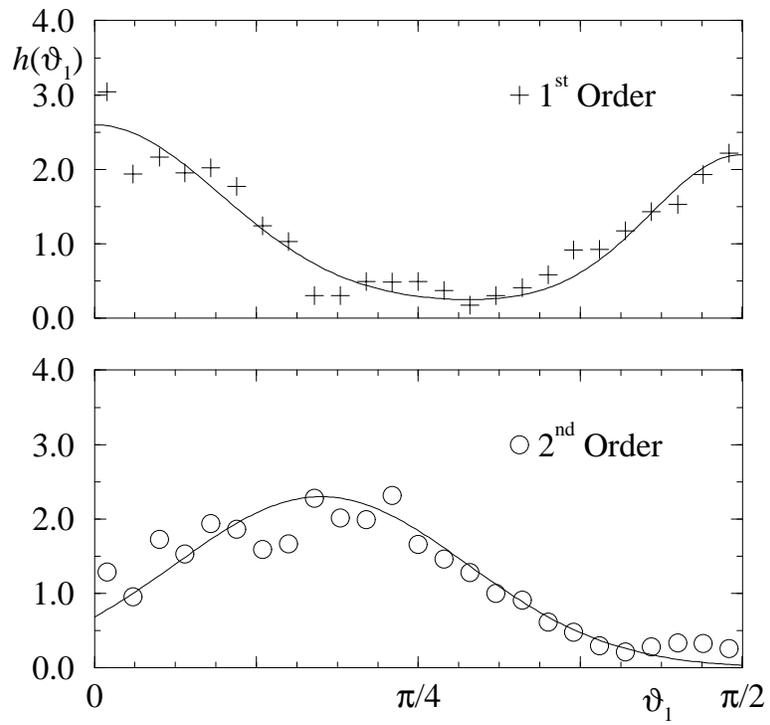

Figure 2